\documentstyle[aps, epsfig]{revtex}

\def \f{\frac}

\newcommand{\be}{\begin{equation}}
\newcommand{\ee}{\end{equation}}
\newcommand{\bea}{\begin{eqnarray}}
\newcommand{\eea}{\end{eqnarray}}
\newcommand{\bean}{\begin{eqnarray*}}
\newcommand{\eean}{\end{eqnarray*}}

\def \ni{\noindent}

\def \mn{\medskip
\ni}

\begin{document}

\title{\Large \bf Boundary terms in the Barrett-Crane spin foam model
\\ and
consistent gluing}

\author{\bf Daniele Oriti}

\address{Department of Applied Mathematics and Theoretical Physics, \\
Centre for Mathematical Sciences, \\ University of Cambridge, \\
Wilberforce Road, Cambridge CB3 0WA, UK \\ and \\ Girton College \\
Cambridge CB3 0JG, UK \\ d.oriti@damtp.cam.ac.uk }

\maketitle

\begin{abstract}
We extend the lattice gauge theory-type derivation of the Barrett-Crane
spin foam model for quantum gravity to other choices of boundary
conditions, resulting in different boundary terms, and re-analyze the
gluing of 4-simplices in this context. This provides a consistency check
of the previous derivation. Moreover we study and discuss some possible
alternatives and variations that can be made to it and the resulting
models. 

\end{abstract}

\section{Introduction}

\ni
Spin foam models \cite{baez,baez2,or}, and the Barrett-Crane one
\cite{BC,BC2} in particular, are promising candidates for the construction
of a quantum theory of gravity from a covariant perspective, implementing
in a purely algebraic fashion the path integral or sum-over-geometries
approach. Different versions of the Barrett-Crane model are present in the
literature \cite{DP-F-K-R,P-R,P-R2,P-R3}, all sharing the same
amplitude for the vertices of the spin foam but using different ampitudes
for its edges, leading to models with different physical properties (in
particular, the perturbative finiteness of one of these versions
\cite{P-R,Perez,CPR} is due to the particular form of the edge
amplitude). 

\mn
In \cite{danruth} a derivation of the Barrett-Crane model was given,
showing how it can be obtained from a discretized BF theory, imposing at
the quantum level (as projectors in the partition function) the analogue
of the constraints that reduce BF theory to gravity in the Plebanski
formulation of GR. The version obtained is the Perez-Rovelli version, thus
shown to come naturally from a discretization of (constrained) BF with
usual methods from lattice gauge theory, being originally derived from a
field theory over a group \cite{P-R}. In order to obtain the exact form of
the edge amplitudes, i.e. the amplitude for the tetrahedra dual to the
edges of the spin foam, the procedure used was to derive first the
expression for the partition function corresponding to a single 4-simplex,
taking into account all the necessary boundary terms, and then to glue
4-simplices along tetrahedra in their boundary, ending up with the
partition function assigned to the  
whole triangulated manifold built up from them. The amplitude to be
assigned to the edges (tetrahedra) results from this gluing procedure only
and does not require any additional input or choice. The fact that it
comes directly from the gluing is to be expected since it should encode
the information describing the (geometric) interaction between
4-simplices. 
The advantage of this lattice gauge theory type of derivation compared
with other existing derivations (being of course strongly related, see in
particular the \lq\lq connection formulation" of field theories over a
group manifold \cite{mike&carlo}), is, in our opinion, that it makes the
link between the Barrett-Crane spin foam model and the classical Plebanski
action \cite{Plebanski,CDJ,DP-F,Rei} more clear, and makes the
analogies  between gravity and lattice gauge theory more explicit.
Moreover, it helps us to understand better the origin and the geometric
meaning of the edge amplitudes in the partition function, and may also
help to clarify the differencies between the various existing versions of
the Barrett-Crane model. On the other hand, this approach has the
shortcoming of being limited to a fixed triangulation of spacetime, while
the field theory over a group allows to sum over all the triangulations,
even if much remains to be understood about this sum.

\mn
In this letter, we extend the derivation of \cite{danruth} to other
choices of boundary conditions, following an analogous study for
3-dimensional gravity \cite{OLough}, obtain the corresponding boundary
terms in the partition function for a single 4-simplex and then apply
again the gluing procedure to get the full partition function for the
triangulated manifold. Apart from giving the correct boundary terms in
this case, this serves as a consistency check for the previous derivation.
In fact it is of course to be expected that the amplitudes for the
elements in the interior of the manifold, the edge amplitudes in
particular, should not be affected by the choice of boundary conditions in
the 4-simplices (having boundaries) whose gluing produces them. The
result is that the derivation in \cite{danruth} is indeed consistent, and
we get again the Perez-Rovelli version of the Barrett-Crane model. We then
examine a few alternatives to the procedure used in \cite{danruth},
exploiting the freedom left by that derivation. In particular, we study
the effect of imposing the projection over the simple representations also
in the boundary terms, since this may (naively) recall the imposition of the
simplicity constraints in the kinetic term in the field theory over the
group manifold, leading to the DePietri-Freidel-Krasnov-Rovelli version of
the Barrett-Crane model \cite{DP-F-K-R}. Instead, this leads in the
present case to several drawbacks, as we discuss, and to a model which is
not the DePietri-Freidel-Krasnov-Rovelli version and it is not consistent,
in the sense specified above, with respect to different choices of
boundary conditions.  Moreover, we study and discuss the model that can be
obtained by not imposing the gauge invariance of the edge amplitude (as
required in \cite{danruth}), since it was mentioned in \cite{hendryk},
explaining why we do not consider it a viable version of the Barrett-Crane
model, and finally the class of models that can be obtained by
imposing the two projections (simplicity and gauge invariance) more
that once. All the calculations in this paper will be performed explicitely
for the Euclidean case, but are valid (or can be easily extended to) in
the Lorentzian case as well, as we will discuss in the following.  

\section{Derivation of the Barrett-Crane spin foam model: constraining and
gluing} \label{sec:der}
Let us now recall the basic elements of the derivation in \cite{danruth}.
The starting point is the expression for the partition function for
$SO(4)$ BF theory discretized on the 2-complex dual to a 4-dimensional
triangulated manifold:
\be
Z_{BF}(SO(4))\,=\,\int_{SO(4)}dg\,\prod_{\sigma}\sum_{J_{\sigma}}\,\Delta_{J_{\sigma}}\,\chi^{J_{\sigma}}(\prod_{e}g_{e})
\ee
where $\sigma$ are the parts of the dual plaquettes associated to each
4-simplex, also called \lq\lq wedges" in the literature
\cite{Reisenberger} (see Fig.1), the sum is over the representations $J$
of $SO(4)$ (given by two half-integer parameters (j,k)) here attached to
each wedge (in such a way that wedges belonging to the same plaquette get
the same representation attached to them),
$\Delta_{J_{\sigma}}=(2j+1)(2k+1)$ is the dimension of the representation,
$\chi^{J}(g)$ is the character of the group element $g$ in the
representation $J$, and the group
variables are associated to the links of the 2-complex, so that for each
wedge there is a group element assigned to it, given by the product of the
group elements associated to the edges of its boundary (see Fig.1). Here
the role of the $B$ field is played by the representations $J$ on the
plaquettes of the 2-complex, and that of the connection by the group
variables on its edges. 

\begin{figure}[tbp]
\centering
\epsfig{figure=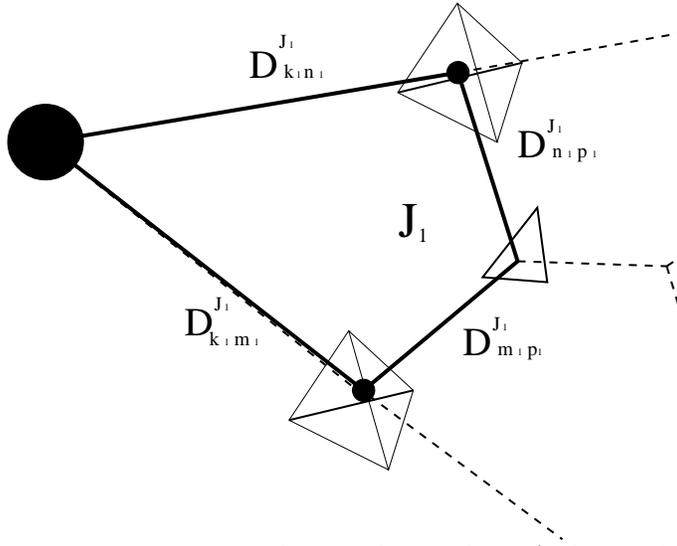, width=9cm}
\caption{Fig.1 - A wedge (the part of a dual face belonging to a
single 4-simplex) with the D-functions for the group elements assigned
to its boundary edges.}
\label{wed}
\end{figure}

An analogous expression can be written in the
Lorentzian case, with $SL(2,C )$ in place of $SO(4)$ and with the
representations labeled by a continuous parameter, $\rho$, and a half-integer
parameter, $n$ (see \cite{BC2,P-R2}), and with ``dimension''
$(n^{2}+\rho^{2})$. Note that this partition function is only formally
defined, since it is divergent, but it will give rise to a well-defined
and convergent expression after the imposition of the Barrett-Crane
constraints.

Now we consider the case of a single 4-simplex, consisting of 5 tetrahedra
(which constitute its boundary), write the character explicitly in terms
of the representation functions of the group elements assigned to each
edge, choosing real representations, and sort the terms in the partition
function per edge, to obtain:

\bea 
\lefteqn{Z_{BF}(SO)=\sum_{J_{\sigma},\{k_{e}\}}\left(\prod_{\sigma}dim_{J_{\sigma}}\right)\prod_{e}\,A_{e}\left(\prod_{\tilde{e}}D\right)  
} \nonumber \\
&=&\sum_{\{J_{\sigma}\},\{k_{e}\}}\left(\prod_{\sigma}\Delta_{J_{\sigma}}\right)\prod_{e}\int_{Spin(4)}dg_{e}\,D_{k_{e1}m_{e1}}^{J_{1}^{e}}D_{k_{e2}m_{e2}}^{J_{2}^{e}}D_{k_{e3}m_{e3}}^{J_{3}^{e}}D_{k_{e4}m_{e4}}^{J_{4}^{e}}\;\left(\prod_{\tilde{e}}D_{il}^{J}\right)\;\;\;\;. 
\eea
There is a group element per edge, so that four representation functions
coming from the four wedges (dual to triangles) incident to it are to be
integrated. Each of these functions has two matrix indices, one referring
to the vertex of the 2-complex (only one vertex since we are considering
only one 4-simplex), and the other, referring to a tetrahedron on the
boundary, contracted with one index of a D-function for an element
attached to (and only to) a link which is exposed on the boundary (see
Fig.1). The other index of each matrix for an exposed link (referring to
a triangle) is contracted with the index coming from the D-function
referring to the same triangle (again, see Fig.1). 

It is crucial to note that the group elements attached to the links
exposed on the boundary for each wedge are {\it not} integrated over,
since we are working with fixed connection on the boundary, a boundary
condition which can be easily shown to not require any additional boundary
term in the classical action (see \cite{OLough} for the 3-dimensional
case).

We then pass from pure BF theory to 4-dimensional gravity imposing the
Plebanski constraints on the $B$ field \cite{Plebanski,CDJ,DP-F,Rei}
directly at the quantum level, i.e. as Barrett-Crane constraints on the
representations $J$ (or $(n,\rho)$ in the Lorentzian case) labelling the
wedges. In turn, this can be done imposing some projections on the edge
amplitude in the partition function we have just described:          
\bea
\lefteqn{A_{e}(GR)\,=\,P_{g}\,P_{h}\,A_{e}(BF)\,=} \nonumber \\
&=&\int_{SO(4)}dg_{1}D_{k_{1}l_{1}}^{J_{1}}(g_{1})D_{k_{2}l_{2}}^{J_{2}}(g_{1})D_{k_{3}l_{3}}^{J_{3}}(g_{1})D_{k_{4}l_{4}}^{J_{4}}(g_{1})
\nonumber \\
&\times&\int_{SO(3)}dh_{1}D_{l_{1}i_{1}}^{J_{1}}(h_{1})\int_{SO(3)}dh_{2}D_{l_{2}i_{2}}^{J_{2}}(h_{2})\int_{SO(3)}dh_{3}D_{l_{3}i_{3}}^{J_{3}}(h_{3})\int_{SO(3)}dh_{4}D_{l_{4}i_{4}}^{J_{4}}(h_{4})
\nonumber \\
&\times&\int_{SO(4)}dg'_{1}\,D_{i_{1}m_{1}}^{J_{1}}(g'_{1})D_{i_{2}m_{2}}^{J_{2}}(g'_{1})D_{i_{3}m_{3}}^{J_{3}}(g'_{1})D_{i_{4}m_{4}}^{J_{4}}(g'_{1}).\label{eq:edam}
\eea

The integrals in the $SO(3)$ $h$ variables impose the simplicity of all
the representations $J$, given by representations of the form $(j,j)$,
while additional integrals over $SO(4)$ restore the gauge invariance of
the edge amplitudes, that, automatic in pure BF theory, is lost after the
imposition of the simplicity constraints. These correspond to the
simplicity and closure constraints in \cite{BC}. In the Lorentzian case
everything works the same way, with $SL(2,C)$ and $SU(2)$
instead of $SO(4)$ and $SO(3)$ \cite{BC2,P-R2}, with the simple
representations given in this case by those labelled only by the
continuous parameter $\rho$.

Note that we are {\it not} imposing any projection in the boundary terms,
so that these are the same as those in pure BF theory. However, the
projection over simple representations in the edge amplitude imposes
automatically the simplicity also of the representations entering in the
D-functions for the exposed edges.

Performing the integrals (we are correcting here a few typos present
in \cite{danruth}) we get:
\be
A_{e}=\frac{1}{\left(\Delta_{J_{1}}\Delta_{J_{2}}\Delta_{J_{3}}\Delta_{J_{4}}\right)^{\frac{1}{2}}}B_{k_{1}k_{2}k_{3}k_{4}}^{J_{1}J_{2}J_{3}J_{4}}B_{m_{1}m_{2}m_{3}m_{4}}^{J_{1}J_{2}J_{3}J_{4}},
\ee 
where the
$B_{k_{1}k_{2}k_{3}k_{4}}^{J_{1}J_{2}J_{3}J_{4}}=\sum_{J}C_{k_{1}k_{2}k_{3}k_{4}}^{J_{1}J_{2}J_{3}J_{4}J}=\sum_{J}\sqrt{\Delta_{J}}C_{k_{1}k_{2}k}^{J_{1}J_{2}J}C_{k_{3}k_{4}k}^{J_{3}J_{4}J}$
are the Barrett-Crane intertwiners, with the $C$s being ordinary $SO(4)$
invariant tensors normalized such that the theta net is equal to one. The
sets of indices of the intertwiners refer to the vertex of the 2-complex
(and are contracted with others coming from the other edges) and to the
boundary tetrahedra (and are contracted with the D-functions for the
exposed edges) (see Fig.2). The factors in the denominator come from
the simplicity projections.

\begin{figure}[tbp]
\centering
\epsfig{figure=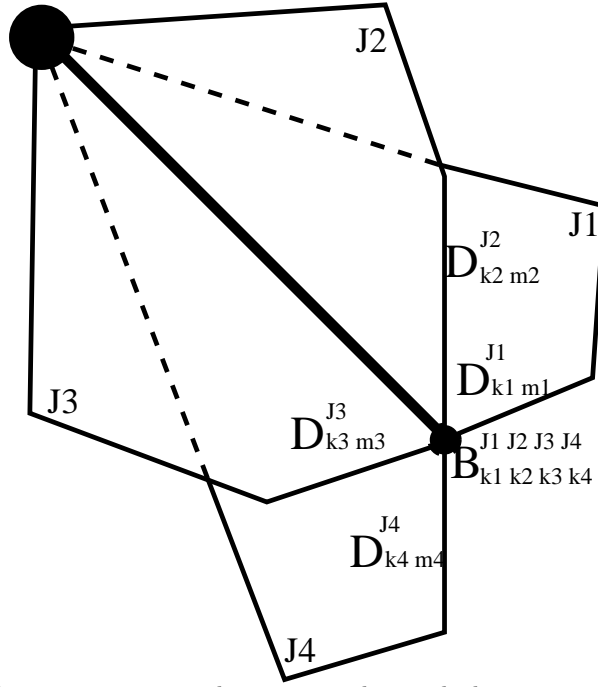, width=8cm}
\caption{Structure of the boundary term corresponding to a single
tetrahedron, i.e. a single dual edge with the 4 wedges incident to it,
and the corresponding 4 exposed edges.}
\label{bou}
\end{figure}

Consequently the partition function for a single 4-simplex is:
\be
Z_{BC}=\sum_{\{J_{f}\},\{k_{e'}\}}\prod_{f}\Delta_{J_{f}}\prod_{e'}\frac{B_{k_{e'1}k_{e'2}k_{e'3}k_{e'4}}^{J_{e'1}J_{e'2}J_{e'3}J_{e'4}}}{\left(\Delta_{J_{e'1}}\Delta_{J_{e'2}}\Delta_{J_{e'3}}\Delta_{J_{e'4}}\right)^{\frac{1}{2}}}\,\prod_{v}{\bf
B}
_{BC}\;\left(\prod_{\tilde{e}}
D(\tilde{g})\right)
\label{eq:4s}
\ee
where the  ${\bf B}_{BC}$ is the Barrett-Crane amplitude for the
4-simplex, and the boundary terms are given by one Barrett-Crane
intertwiner for each tetrahedron on the boundary, and one D-function for
each group element on each of the exposed edges, contacted with the
intertwiner to form a group invariant (see Fig.2), plus a \lq\lq
regularizing" factor in the denominator.

The gluing of 4-simplices is now simply done by multiplying the partition
functions for the individual 4-simplices, and integrating over the group
variables that are not anymore on the boundary of the manifold, and
required to be equal in the two 4-simplices, again because we are working
with fixed connection on the boundary, so that the boundary data of the
two 4-simplices being glued have to agree.

These group variables appear only in two exposed edges each, and the
orthogonality between D-functions forces the representations corresponding
to the two wedges to be equal:

\be
\int_{SO(4)}dg
D^{J}_{kl}(g)\,D^{J'}_{mn}(g)\,=\,\f{1}{\Delta_{J}}\delta_{km}\delta_{ln}\delta_{JJ'};
\label{eq:ort} \ee    
moreover, the factors $\f{1}{\Delta_{J}}$ compensate for having two wedges
corresponding to the same triangle, so that to each plaquette of the dual
complex, or triangle of the triangulation, corresponds still only a factor
$\Delta_{J}$ in the partition function. Finally, the equality of the
matrix indices in the previous relation forces the Barrett-Crane
intertwiners corresponding to the same shared tetrahedron to be fully
contracted, so that the resulting amplitude for it (taking properly into
account the normalization chosen above) is:
\be
A_{e}\,=\,\frac{B_{k_{1}k_{2}k_{3}k_{4}}^{J_{1}J_{2}J_{3}J_{4}}B_{k_{1}k_{2}k_{3}k_{4}}^{J_{1}J_{2}J_{3}J_{4}}}{\Delta_{J_{1}}\Delta_{J_{2}}\Delta_{J_{3}}\Delta_{J_{4}}}=
\frac{\Delta_{1234}}{\Delta_{J_{1}}\Delta_{J_{2}}\Delta_{J_{3}}\Delta_{J_{4}}}
\label{eq:ampl}
\ee
where $\Delta_{1234}$ is the dimension of the space of intertwiners
between the representations $J_{1},..,J_{4}$, i.e. the number of possible
intertwiners between these representations. Note that this is also the
number of possible quantum tetrahedra for given values of their triangle
areas, so it is really the most natural statistical weight for them in
the partition function.

Note that the gluing is not trivial, in the sense that the end result is
not just the product of pre-existing factors, but includes something
resulting from the gluing itself (the factor $\Delta_{1234}$).

In the end, the partition function we find for a general manifold with
boundary, with fixed connection on the boundary, is:

\be
Z_{BC}=\sum_{\{j_{f}\},\{k_{e'}\}}\prod_{f}\Delta_{j_{f}}\prod_{e'}\frac{B_{k_{e'1}k_{e'2}k_{e'3}k_{e'4}}^{j_{e'1}j_{e'2}j_{e'3}j_{e'4}}}{\left(\Delta_{j_{e'1}}\Delta_{j_{e'2}}\Delta_{j_{e'3}}\Delta_{j_{e'4}}\right)^{\frac{1}{2}}}\prod_{e}\frac{\Delta_{1234}}{\Delta_{j_{e1}}\Delta_{j_{e2}}\Delta_{j_{e3}}\Delta_{j_{e4}}}\prod_{v}{\bf
B}_{BC}
\;\left( \prod_{\tilde{e}}D(g_{\tilde{e}}\right)\label{eq:final}
\ee
where the $\{e'\}$ and the $\{e\}$ are the sets of boundary (incident to
it) and interior edges of the spin foam, respectively, while the
$\tilde{e}$ are the remaining exposed edges, where the boundary connection
data are located. The partition function is then a function of the
connection, i.e. of the group elements on the exposed edges.
This is the Perez-Rovelli version of the Barrett-Crane model, with the
appropriate boundary terms.

One can proceed analogously in the Lorentzian case, using the integral
representation of the Barrett-Crane intertwiners (the resulting expression
is of course more complicated, but with the same structure), and their
formula for the evaluation of relativistic (simple) spin networks. All
the passages above, in fact, amount to the evaluation of spin networks,
which were proven to evaluate to a finite number, so the procedure above
can be carried through similarly and sensibly.
   
We see that, starting from a ill-defined BF partition function, the
imposition of the constraints has made the resulting partition function
for gravity finite both in the Euclidean and Lorentzian cases
\cite{P-R,Perez,CPR}.

\section{Fixing the boundary metric} \label{sec:bound}
Let us now study the case in which we choose to fix the B field on the
boundary (i.e. by the metric field), and let us analyse first the
classical action.

We note here that the partition function we will obtain in this section,
being a function of the representations $J$ (or $\rho$) of the group
$SO(4)$ (or $SL(2,C )$ assigned to the boundary, and representing
the B (metric) field, can be thought of as the Fourier transform
\cite{dani2} of the one we ended up with in the previous section, being
instead function of the group elements, representing the connection field.

The $so(4)$ Plebanski action is:

\be
S\,=\,\int_{\mathcal{M}}B\wedge F\,+\,\frac{1}{2}\phi\,B\,\wedge\,B
\ee
so that its variation si simply given by:
\be
\delta S\,=\,\int_{\mathcal{M}}\delta B\wedge \left(F\,+\,\phi\,B\right) +
\delta A \wedge \left(dB\,+\,A\wedge B\,+\,B\wedge A\right) \,-\,
\int_{\partial \mathcal{M}} B\wedge\delta A,
\ee
and we see that fixing the connection on the boundary does not require any
additional boundary term to give a well-defined variation, i.e. the field
equations resulting from it are not affected by the presence of a
boundary.

On the other hand, if we choose to fix the B field on the boundary, we
need to introduce a boundary term in the action:
\be
S\,=\,\int_{\mathcal{M}}B\wedge
F\,+\,\frac{1}{2}\phi\,B\,\wedge\,B\,+\,\int_{\partial\mathcal{M}}
B\wedge A
\ee
so that the variation leads to:
\be
\delta S\,=\,\int_{\mathcal{M}}\delta B\wedge \left(F\,+\,\phi\,B\right) +
\delta A \wedge \left(dB\,+\,A\wedge B\,+\,B\wedge A\right) \,+\,
\int_{\partial \mathcal{M}} \delta B\wedge A,
\ee
and to the usual equations of motion.  

Now we want to find what changes in the partition function for a single
4-simplex if we decide to fix the B field on the boundary, and then to
study how the gluing proceeds in this case. 

The additional term in the partition function resulting from the
additional term in the action is $\exp{\int_{\partial\mathcal{M}}B\wedge
A}$. We have to discretize it, expressing it in terms of representations
$J$ and group elements $g$ on the boundary, and then multiply it into the
existing amplitude. The connection terms on the boundary are then to be
integrated out, since they are not held fixed anymore, while the sums over
the representations have to be performed only on the bulk ones, i.e. only
on the representations labelling the triangles in the interior of the
manifold (none in the case of a single 4-simplex).
A natural discretization \cite{OLough} for the additional term is:
\be
\exp{\int_{\partial\mathcal{M}}B\wedge
A}\,=\,\prod_{\tilde{e}}\chi^{J}(g_{\tilde{e}})\,=\,\prod_{\tilde{e}}D_{kk}^{J}(g_{\tilde{e}})
\label{eq:bt}
\ee   
where the representation $J$ is the one assigned to a wedge with edges
exposed on the boundary, and $g_{\tilde{e}}$ is actually the product
$g_{1}g_{2}$ of the group elements assigned to the two edges exposed on the
boundary, and the product runs over the exposed parts of the wedges. 

We multiply the partition function (~\ref{eq:4s}) by this extra term, and
integrate over the group elements, simultaneously dropping the sum over
the representations, since all the wedges are on the boundary, and thus
all the representations are fixed. 

Using again the orthogonality of the D-functions, eq.~\ref{eq:ort}, the
result is the following:

\be
Z_{BC}=\,\prod_{f}\Delta_{J_{f}}\prod_{e'}\frac{B_{k_{e'1}k_{e'2}k_{e'3}k_{e'4}}^{J_{e'1}J_{e'2}J_{e'3}J_{e'4}}}{\Delta_{J_{e'1}}\Delta_{J_{e'2}}\Delta_{J_{e'3}}\Delta_{J_{e'4}}}\prod_{v}{\bf
B}_{BC}
\label{eq:4s'}
\ee

where also the $k$ indices are fixed by the only constraint (coming
again from the integration over the group above) that the $k$s appearing
in different Barrett-Crane intertwiners but referring to the same triangle
must be equal. Of course we see that the partition function is now a
function of the representations $J$ on the boundary and of their
projections. The different power in the denominator of the boundary terms
is necessary to have consistency in the gluing procedure, as we will see.
Also, note that we did not impose any projection over the simple
representations in the boundary terms, i.e. in the D-functions coming from
the additional boundary term in the action, since we decided not to impose
it in the D-functions for the exposed edges in (~\ref{eq:4s}) above. We will
analyse the alternatives to this choice in the next section.

Now we proceed with the gluing of 4-simplices. The different 4-simplices
being glued have to share the same boundary data for the common
tetrahedron, i.e. the representations $J$ and the projections $k$ in the
Barrett-Crane intertwiner referring to it have to agree. The gluing is
performed again multiplying the partition functions for the two
4-simplices and summing over the $k$s, because they are now attached to a
tetrahedron in the interior of the manifold. In this way the Barrett-Crane
intertwiners corresponding to the same tetrahedron in the two 4-simplices
get contracted with each other, and they give again a factor
$\Delta_{1234}$ as before. The factors in the denominators of the
(ex-)boundary terms are multiplied to give a factor
$1/(\Delta_{J_{1}}\Delta_{J_{2}}\Delta_{J_{3}}\Delta_{J_{4}})^{2}$, but
since we have a factor of $\Delta_{J_{i}}$ for each wedge and for each
4-simplex, the factor in the denominator of the amplitude for the interior
tetrahedron is again
$1/(\Delta_{J_{1}}\Delta_{J_{2}}\Delta_{J_{3}}\Delta_{J_{4}})$. 

In the end the partition function for a generic manifold with boundary,
with the boundary condition being that the metric field is fixed on it,
is:
\be
Z_{BC}=\sum_{\{J_{f}\}}\prod_{f}\Delta_{J_{f}}\prod_{e'}\frac{B_{k_{e'1}k_{e'2}k_{e'3}k_{e'4}}^{J_{e'1}J_{e'2}J_{e'3}J_{e'4}}}{\Delta_{J_{e'1}}\Delta_{J_{e'2}}\Delta_{J_{e'3}}\Delta_{J_{e'4}}}\prod_{e}\frac{\Delta_{1234}}{\Delta_{J_{e1}}\Delta_{J_{e2}}\Delta_{J_{e3}}\Delta_{J_{e4}}}\prod_{v}
{\bf B}_{BC}.
\ee

It is understood that the sum over the representations $J$s is only over
those labelling wedges (i.e. faces) in the interior of the manifold, the
others being fixed.

We recall that this can be understood as the probabiblity amplitude for
the boundary data, the representations of $SO(4)$ (or $SL(2,C)$)
in this case or the $SO(4)$ group elements as in the previous section, in
the Hartle-Hawking vacuum. If the boundary data are instead divided into
two different sets, then the partition function represents the transition
amplitude from the data in one set to those in the other.
The Lorentzian case, again, goes similarly, with the same result.

We see that, apart from the boundary terms, we ended up again with the
Perez-Rovelli version of the Barrett-Crane model. This was to be expected,
since the bulk partition function should not be affected by the boundary
conditions we have chosen for the single 4-simplices before performing the
gluing, but the fact that this is indeed the case represents a good
consistency check for the whole procedure we used to obtain the
Barrett-Crane model from a discretized BF theory. 

\section{Exploring alternatives}
Let us now go on to explore the alternatives to the procedure we have just
used, to see whether there are other consistent procedures giving
different results, i.e. different versions of the Barrett-Crane model.      
In particular we would like to see, for example, whether there is any variation of the
procedure used above resulting in the DePietri-Freidel-Krasnov-Rovelli
version of the Barrett-Crane model \cite{DP-F-K-R}, i.e. the other version
that can be derived from a field theory over a group manifold.
Again, the analogous calculations in the Lorentzian case go through
similarly. 

We have seen in section ~\ref{sec:der} that two choices were involved in
the derivation we performed: the way we imposed the constraints, with one
projection imposing simplicity of the representations and the other
imposing the invariance under the group of the edge amplitude, and the way
we treated the D-functions for the exposed edges, i.e. without imposing
any constraints on them. We will now consider alternatives to these
choices, starting from the last one. A few other alternatives to the first
were considered in \cite{danruth}.

\subsection{Projections on the exposed edges} 
We then first leave the edge amplitude (~\ref{eq:edam}) as it is, and look
for a way to insert an integral over the $SO(3)$ subgroup in the boundary
representation functions.
The idea of imposing the simplicity projections in the D-functions for the
exposed edges may (naively) resemble the imposition of them in the kinetic term in
the action for the field theory over a group, leading to the
DePietri-Freidel-Krasnov-Rovelli version of the Barrett-Crane model
\cite{DP-F-K-R}, since in both cases there are precisely 4 of them for
each tetrahedron, and they represent the boundary data to be transmitted
across the 4-simplices (in the connection representation). Anyway, this is
not the case, as we are going to show.

There are two possible ways of imposing the projections, corresponding to
the two possibilities of multipliying the arguments of the D-functions by
an $SO(3)$ element from the left or from the right, corresponding to
projecting over an $SO(3)$ invariant vector the indices of the D-functions
referring to the tetrahedra or those referring to the triangles (see
figure3), then integrating over the subgroup as in (~\ref{eq:edam}), having
for each boundary term: 
\bea
\lefteqn{B_{k_{1}k_{2}k_{3}k_{4}}^{J_{1}J_{2}J_{3}J_{4}}D^{J_{1}}_{k_{1}m_{1}}(g_{1})D^{J_{2}}_{k_{2}m_{2}}(g_{2})D^{J_{3}}_{k_{3}m_{3}}(g_{3})D^{J_{4}}_{k_{4}m_{4}}(g_{4})\rightarrow}
\nonumber \\ &\rightarrow&
B_{k_{1}k_{2}k_{3}k_{4}}^{J_{1}J_{2}J_{3}J_{4}}D^{J_{1}}_{k_{1}l_{1}}(g_{1})D^{J_{2}}_{k_{2}l_{2}}(g_{2})D^{J_{3}}_{k_{3}l_{3}}(g_{3})D^{J_{4}}_{k_{4}l_{4}}(g_{4})w^{J_{1}}_{l_{1}}w^{J_{2}}_{l_{2}}w^{J_{3}}_{l_{3}}w^{J_{4}}_{l_{4}}w^{J_{1}}_{m_{1}}w^{J_{2}}_{m_{2}}w^{J_{3}}_{m_{3}}w^{J_{4}}_{m_{4}}
\label{eq:1} \eea

or:

\bea
\lefteqn{B_{k_{1}k_{2}k_{3}k_{4}}^{J_{1}J_{2}J_{3}J_{4}}D^{J_{1}}_{k_{1}m_{1}}(g_{1})D^{J_{2}}_{k_{2}m_{2}}(g_{2})D^{J_{3}}_{k_{3}m_{3}}(g_{3})D^{J_{4}}_{k_{4}m_{4}}(g_{4})\rightarrow}
\nonumber \\ &\rightarrow&
B_{k_{1}k_{2}k_{3}k_{4}}^{J_{1}J_{2}J_{3}J_{4}}w^{J_{1}}_{k_{1}}w^{J_{2}}_{k_{2}}w^{J_{3}}_{k_{3}}w^{J_{4}}_{k_{4}}w^{J_{1}}_{l_{1}}w^{J_{2}}_{l_{2}}w^{J_{3}}_{l_{3}}w^{J_{4}}_{l_{4}}D^{J_{1}}_{l_{1}m_{1}}(g_{1})D^{J_{2}}_{l_{2}m_{2}}(g_{2})D^{J_{3}}_{l_{3}m_{3}}(g_{3})D^{J_{4}}_{l_{4}m_{4}}(g_{4})
\label{eq:2}
\eea

where in the first case the second set of invariant vectors is contracted
with one coming from another boundary term, giving in the end no
contribution to the amplitude, while in the second case there is a
contraction between the Barrett-Crane intertwiners and these vectors,
giving a different power in the denominator in (~\ref{eq:4s}), and the
disappearence of the intertwiners from the amplitude.

Let us discuss the first case. The effect of the projection is to break
the gauge invariance of the amplitude for a 4-simplex, and to decouple the
different tetrahedra on the boundary. In fact the amplitude for a
4-simplex is then:
 \be
Z=\sum_{\{J_{f}\},\{k_{e'}\}}\prod_{f}\Delta_{J_{f}}\prod_{e}\frac{B_{k_{e1}k_{e2}k_{e3}k_{e4}}^{J_{e1}J_{e2}J_{e3}J_{e4}}}{\left(\Delta_{J_{e1}}\Delta_{J_{e2}}\Delta_{J_{e3}}\Delta_{J_{e4}}\right)^{\frac{1}{2}}}D^{J_{e1}}_{k_{e1}l_{e1}}(g_{e1})...D^{J_{e4}}_{k_{e4}l_{e4}}(g_{e4})w^{J_{e1}}_{l_{e1}}w^{J_{e2}}_{l_{e2}}w^{J_{e3}}_{l_{e3}}w^{J_{e4}}_{l_{e4}}w^{J_{e1}}_{m_{e1}}w^{J_{e2}}_{m_{e2}}w^{J_{e3}}_{m_{e3}}w^{J_{e4}}_{m_{e4}}\prod_{v}{\bf
B}_{BC}
\label{eq:4s3}
\ee
which is not gauge invariant but only gauge covariant. 

This would be enough for discarding this variation of the procedure used
above as not viable. Nevertheless, this does not lead to any apparent
problem when we proceed with the gluing as we did previously. In fact, as
it can be easily verified, the additional invariant vectors $w$ do not
contribute to the gluing, when the connection is held fixed at the
boundary, and the result is again the ordinary Perez-Rovelli version of
the Barrett-Crane model. The edge amplitude, i.e. the amplitude for the
tetrahedra in the interior of the manifold, is again given by
(~\ref{eq:ampl}).
However, the inconsistency appears when we apply the \lq\lq consistency
check" used previously, i.e. when we study the gluing with different
boundary conditions. In fact, when the field $B$ is held fixed at the
boundary, we have to multiply again the partition function (~\ref{eq:4s3})
by the additional boundary terms ~\ref{eq:bt}, this time imposing the
simplicity projections here as well as in (~\ref{eq:1}). 
The resulting 4-simplex amplitude is:     
\be
Z=\,\prod_{f}\Delta_{J_{f}}\prod_{e'}\frac{B_{k_{e'1}k_{e'2}k_{e'3}k_{e'4}}^{J_{e'1}J_{e'2}J_{e'3}J_{e'4}}}{\left(\Delta_{J_{e'1}}\Delta_{J_{e'2}}\Delta_{J_{e'3}}\Delta_{J_{e'4}}\right)^{\frac{3}{2}}}\prod_{v}{\bf
B}_{BC}
\label{eq:4s3'}
\ee
and the gluing results in an edge amplitude for the interior tetrahedra: 
\be
A_{e}=\frac{\Delta_{1234}}{\left(\Delta_{J_{1}}\Delta_{J_{2}}\Delta_{J_{3}}\Delta_{J_{4}}\right)^{2}}. 
\ee

This proves that this model is not consistent, since the result is
different for different boundary conditions, and shows also that, as we
said above, the \lq\lq consistency check" is not trivially satisfied by
every model. 
 
Considering now the second variation (~\ref{eq:2}), we see that imposing the
simplicity constraint this way gives the same result as if we had imposed
it directly in the edge amplitude (~\ref{eq:edam}), having
$A_{e}(GR)=P_{h}P_{g}P_{h}A_{e}(BF)$. This, however, breaks the gauge
invariance of the edge amplitude, for which we were aiming when we imposed
the additional projection $P_{g}$. In turn this results into a breaking of
the gauge invariance of the 4-simplex amplitude. Because of this we do
not
explore any further this variation, but rather study directly the simpler
case in which we do not impose the projection $P_{g}$ at all in the edge
amplitude. Then we will give more reasons for imposing it. 

\subsection{Imposing the projections differently}
We then study the model obtained dropping the projection $P_{g}$ in
(~\ref{eq:edam}), and not imposing any additional simplicity projection on
the D-functions for the exposed edges, since we have just seen that this
would lead to inconsistencies (more precisely, projecting the indices
referring to the triangles would lead to inconsistencies, while projecting
those referring to the tetrahedra would give exactly the same result as
not projecting at all, as can be verified).

The edge amplitude replacing (~\ref{eq:edam}) is then:
\be
A_{e}=\frac{B_{k_{1}k_{2}k_{3}k_{4}}^{J_{1}J_{2}J_{3}J_{4}}}{\left(\Delta_{J_{1}}\Delta_{J_{2}}\Delta_{J_{3}}\Delta_{J_{4}}\right)^{\frac{1}{4}}}\,w^{J_{1}}_{m_{1}}w^{J_{2}}_{m_{2}}w^{J_{3}}_{m_{3}}w^{J_{4}}_{m_{4}}
\ee
and the partition function for a single 4-simplex is:
\be
Z_{BC}=\sum_{\{J_{f}\},\{k_{e'}\}}\prod_{f}\Delta_{J_{f}}\prod_{e'}\frac{w^{J_{1}}_{m_{1}}w^{J_{2}}_{m_{2}}w^{J_{3}}_{m_{3}}w^{J_{4}}_{m_{4}}}{\left(\Delta_{J_{e'1}}\Delta_{J_{e'2}}\Delta_{J_{e'3}}\Delta_{J_{e'4}}\right)^{\frac{1}{4}}}\prod_{v}{\bf
B}_{BC}\;\left(\prod_{\tilde{e}}D\right), 
\ee
where the D-functions for the exposed edges are contracted not with the
Barrett-Crane intertwiners but with the $SO(3)$ invariant vectors
$w^{J}_{m}$. Consequently the partition function itself is not an
invariant under the group. However, let us go a bit further to see which
model results from the gluing.
Proceeding to the usual gluing, the resulting edge amplitude for interior
tetrahedra is simply:
\be
\frac{1}{\sqrt{\Delta_{J_{e'1}}\Delta_{J_{e'2}}\Delta_{J_{e'3}}\Delta_{J_{e'4}}}} 
\ee
and the gluing itself looks rather trivial in the sense that in the end it
just gives a multiplication of pre-existing factors, with nothing new
arising from it. The gluing performed starting from the partition function
with the other boundary conditions gives the same result, again only if we
do not project the D-functions for the exposed edges. 

This is the \lq\lq factorized" edge amplitude considered in
\cite{hendryk}, and singled out by the requirement that the passage from
$SO(4)$ BF theory to gravity is given by a pure projection operator
(how the dual or connection picture changes for the Perez-Rovelli
version is shown in \cite{dani2}).
Indeed, we have just seen how this model is obtained using only the
implicity projection, and dropping the $P_{g}$, which is responsible for
making the combined operator $P_{g}P_{h}$ not a projector
($P_{g}P_{h}P_{g}P_{h}\neq P_{g}P_{h}$). 

On the other hand, the additional projection $P_{g}$ introduces an
additional coupling of the representations for the four triangles
forming a tetrahedron. This coupling
can be understood algebraically directly from the way the $P_{g}$ operator
acts, since it involves all the four wedges incident to the same edge (see
equation (~\ref{eq:edam})), or recalling that the gauge invariance of the edge
amplitude (corresponding to the tetrahedra of the simplicial manifold)
admits a natural interpretation as the closure constraint for the
bivectors $B$ in terms of which we quantize both BF theory and gravity in
the Plebanski formulation. This is the constraint that the bivectors
assigned to the triangles of the tetrahedron, forced
to be simple bivectors because of the simplicity constraint $P_{h}$,
sum to zero. Thus we can argue more
geometrically for the necessity of the $P_{g}$ projection saying that
the model has to describe the geometric nature of the
triangles, but also the way they are \lq\lq coupled" to form \lq\lq
collective structures", like tetrahedra. Not imposing the $P_{g}$
projection results in a theory of not enough coupled triangles. For
this reason we do not consider this as a
viable version of the Barrett-Crane model.

But if the $P_{g}$ projector is necessary, then the procedure of
sections ~\ref{sec:der}~\ref{sec:bound}, giving the Perez-Rovelli version of the
Barrett-Crane model, can be seen as the minimal, and most natural, way of
constraining BF-theory to get a quantum gravity model. At the same time,
exactly because combining the projectors $P_{h}$ and $P_{g}$ does not give
a projector operator, \lq\lq non-minimal" models, sharing the same
symmetries of the Perez-Rovelli version, and implementing as well the
Barrett-Crane constraints, but possibly physically different from it, can
be easily constructed, imposing the two projectors more than once. It is
easy to verify that, both starting from the partition function for a
single 4-simplex with fixed boundary connection or with the B field fixed
instead, imposing the combined $P_{g}P_{h}$ operator $n$ times ($n\geq
1$), the usual
gluing procedure will result in an amplitude for the interior tetrahedra:
\be
A_{e}\,=\,\frac{\Delta_{1234}^{2n-1}}{\left(\Delta_{J_{1}}\Delta_{J_{2}}\Delta_{J_{3}}\Delta_{J_{4}}\right)^{n}}.
\ee    
Of course, the same kind of model could be obtained from a field theory
over a group, with the usual technology. However, the physical
significance of this variation is unclear (apart from the stronger
convergence of the partition function, which is quite apparent).

To conclude, let us comment on the De Pietri-Freidel-Krasnov-Rovelli
version of the Barrett-Crane model. It seems that there is no natural (or
simple) variation of the procedure we used leading to this version of the
Barrett-Crane state sum, as we have seen. In other words, starting from
the partition function for BF theory, it appears to be no simple way to
impose the Barrett-Crane constraints at the quantum level, by means of
projector operators as we did, and to obtain a model with an amplitude for
the interior tetrahedra of the type:
\be
A_{e}\,=\,\frac{1}{\Delta_{1234}} \label{eq:dpfkredam}
\ee     
as in \cite{DP-F-K-R}. Roughly, the reason can be understood as follows:
for each edge, the $P_{h}$ projection has the effect of giving a factor
involving the product of the dimensions of the representations in the
denominator, and of course of restricting the allowed representations to
the simple ones, while the $P_{g}$ projector is responsible for having a
Barrett-Crane intertwiner for the boundary tetrahedra, which in turn
produces the factor $\Delta_{1234}$ after the gluing. The imposition of
more of these projections in the non-minimal models can only change the
power with which these same elements appear in the final partition
function, as we said. So it seems that the imposition of these projectors
can not create a factor like $\Delta_{1234}$ in the denominator, which, if
wanted, has apparently to be inserted by hand from the beginning. The
un-naturality of this version of the Barrett-Crane model from this point
of view can probably be understood noting that in the original field
theory over group formulation \cite{DP-F-K-R} the imposition of the
operator $P_{h}$ in the kinetic term of the action, giving a kinetic
operator that is not a projector anymore, makes the coordinate space (or
\lq\lq connection" \cite{mike&carlo}) formulation of the partition
function highly complicated, and this formulation is the closest
analogue of our lattice-gauge-theory-type of derivation. However, an intriguing logical
possibility that we think deserves further study is that the edge
amplitude (~\ref{eq:dpfkredam}) may be ``expanded in powers of
$\Delta_{1234}$'', so that it may arise from a (probably asymptotic) series in which the
n-th term results from imposing the $P_{g}P_{h}$ operator $n$ times with the
result shown above. More generally, many different models can be
constructed (consistently with different boundary conditions) in this way (combining models with different powers of $P_{g}P_{h}$), all based on the simple representations of
$SO(4)$ or $SL(2,C)$, having the same fundamental symmetries,
and the Perez-Rovelli version of the Barrett-Crane model as the
``lowest order'' term, with the other orders as ``corrections'' to
it, even if interesting models on their own right. This possibility will be investigated in the future.   

\section{Conclusions}
We have thus shown how the Barrett-Crane spin foam model for quantum
gravity (in the Perez-Rovelli version) can be obtained with a lattice
gauge theory type of derivation, with the appropriate boundary terms
corresponding to fixing the B (metric) field on the boundary of the
manifold. We stress that the correct treatment of the boundary terms and
their precise description will be necessary for any concrete application
of the spin foam model, like for example the calculation of transition
amplitudes between quantum gravity states \cite{P-R4}, or the study of
black hole physics (e.g computing black hole entropy) \cite{OLough,SK}.
We have also described how the gluing between 4-simplices has to be
carried out in this context. The result is consistent with the one
obtained in \cite{danruth} fixing the connection field instead. We also
explored several variations of this derivation, including one
resulting in a class of ``non-minimal'' models that may turn out to be
useful in the future. As a result, the
Perez-Rovelli version of the Barrett-Crane model appears to be the simplest consistent outcome of constraining BF theory with a procedure of the
kind we used.   

\section*{Acknowledgements}
The author is much grateful to E. R. Livine, H. Pfeiffer and
R. M. Williams for many discussions and suggestions.

\end{document}